\documentclass[aps,reprint, twocolumn,nofootinbib,showpacs,floats]{revtex4}

\usepackage{color}
\usepackage[normalem]{ulem}
\usepackage{amsmath}
\usepackage{enumerate}
\usepackage{amsfonts}
\usepackage{epsfig}
\usepackage{graphicx}
\usepackage{bm}
\usepackage{dcolumn}

\newcommand{\be}{\begin{equation}}
\newcommand{\ee}{\end{equation}}
\newcommand{\bea}{\begin{eqnarray}}
\newcommand{\eea}{\end{eqnarray}}

\newcommand{\bln}{\begin{align}}
\newcommand{\eln}{\end{align}}
\newcommand{\bst}{\begin{split}}
\newcommand{\est}{\end{split}}
\newcommand{\bi}{\begin{itemize}}
\newcommand{\ei}{\end{itemize}}
\newcommand{\ben}{\begin{enumerate}}
\newcommand{\een}{\end{enumerate}}

\def\eeq{\end{equation}}

\begin{document}

\title {Disorder in Gauge/Gravity Duality, Pole Spectrum Statistics
\\and\\
Random Matrix Theory }


\author{Omid Saremi}
\bigskip
\affiliation{Berkeley Center for Theoretical Physics and Department of Physics, University of California, Berkeley, CA 94720-7300\\
and \\Theoretical Physics Group, Lawrence Berkeley National Laboratory, Berkeley, CA  94720-8162}  
\begin{abstract}
In condensed-matter, level statistics has long been used to characterize the phases of a disordered system. We provide evidence within the context of a simple model that in a disordered large-N gauge theory with a gravity dual, there exist phases where the nearest neighbor spacing distribution of the unfolded pole spectra of generic two-point correlators is Poisson. This closely resembles the localized phase of the Anderson Hamiltonian. We perform two tests on our statistical hypothesis. One is based on a statistic defined in the context of Random Matrix Theory, the so-called $\overline{\Delta_3}$, or spectral rigidity, proposed by Dyson and Mehta. The second is a $\chi$-squared test. In our model, the results of both tests are consistent with the hypothesis that the pole spectra of two-point functions can be at least in two distinct phases; first a regular sequence and second a completely uncorrelated sequence with a Poisson nearest neighbor spacing distribution. 
\end{abstract}
\maketitle
\newpage

Real-world condensed-matter systems are always impure; realistic samples are often described by non-translationally invariant interactions, about which only limited statistical information is available. As usual, perturbation theory has a limited range of applicability and answering questions concerning the phase structure of a disordered system generally requires knowledge of non-perturbative dynamics. A prime example is the celebrated Anderson Localization (AL) phenomenon \cite{Anderson}, which, in sufficiently high dimensions, is entirely inaccessible by perturbation theory. 

In recent years, the Anti-de Sitter/Conformal Field Theory (AdS/CFT) correspondence \cite{Malda} or more generally gauge/gravity duality has taught us a great deal about certain strongly correlated large-N gauge theories\footnote{Here ``N'' denotes the number of colors. In this work, we will be working only in the gravity limit, where N is large.}, by offering a dual gravitational perspective. It has provided us with a computational edge, in uncomfortable regions of parameter space of field theories, where the conventional perturbative techniques fail. Over the past few years, understanding quantum critical points in various condensed matter systems using gauge-gravity duality \cite{Sean}, \cite{John}, nicknamed AdS/Condensed-Matter Theory (AdS/CMT), has emerged as a new avenue of research, where the correspondence, at least qualitatively, has proven insightful. In this approach one writes down phenomenological gravitational dynamics which are expected to capture some aspects of the condensed matter system under study. 
Due to technical challenges, disordered systems have not received much attention.  In \cite{SeanChris}, the effects of perturbative disorder on transport phenomena were first pioneered. Authors of \cite{Adams1}, \cite{Adams2} studied renormalization group properties of disorder in holographic settings. Holography and structural disorder/glassiness were studied in \cite{Dio1}, \cite{Dio2}. See also \cite{Taka} for a replica-trick formulation of disorder-averages in AdS/CFT. 

In this paper we seek to address the following pressing question: what constitutes a ``generic'' hallmark of a disorder-induced phase transition in a system possessing a gravity dual?\footnote{It would be advantageous if the existence of the order-parameter does not rely on perturbation theory.} A disorder-induced phase transition of central importance is the AL transition \cite{Anderson}. In dealing with disordered systems in condensed-matter, one approach has been to study statistical properties of the levels, i.e, the discrete energy spectrum of a given disordered system. The particular statistic of interest is the so-called nearest neighbor spacing distribution (NNSD), defined in the context of Random Matrix Theory (RMT). Level spacing distribution is a reliable order-parameter to distinguish localized from delocalized phases, when a disordered system undergoes an AL transition  \cite{Ault}, \cite{Imry}, \cite{Izra}. For instance, the disorder-driven Metal-Insulator Transition, induced by AL, is marked by a non-perturbative change in the NNSD of levels from Gaussian Orthogonal Ensemble (GOE) universality class\footnote{Depending on microscopic symmetries of the Hamiltonian, the GOE here must be replaced by one of the other Gaussian Ensembles.} in the delocalized regime to \emph{Poisson} in the localized phase \cite{Shk}.

In this paper we demonstrate, in the framework of a simple model, the existence of a similar phenomenon, where the ``condensed matter system'' is a field theory with a gravity dual and non-translationally invariant interactions. We work with quenched disorder only.

At large-N, even at finite volume, level statistics is not a good and/or convenient observable.
In this paper, we put forward an alternative proposal, suggesting that for a large-N gauge theory a candidate order-parameter is the NNSD of the pole spectra of generic two-point correlators. It is important to note that the discrete pole spectrum does not represent the full state content of a many-body system. We consider the case where the pole spectrum consists of at least one discrete pole sequence, containing large number of poles. In the toy model discussed in the present work, there are infinitely many discrete poles all lying on the real axis. 

In a generic disordered system, there exists a set of operators $\{\mathcal{O}_{\Phi_{I}}\}$ with position-dependent source terms, deforming the theory. Whenever a gravity dual is available, the operators $\{\mathcal{O}_{\Phi_{I}}\}$ will be dual to bulk fluctuations $\{\Phi_{I}\}$ with prescribed position-dependent boundary conditions. In general, the back-reacted bulk geometry will have no Killing symmetries. Using AdS/CFT, we compute the real-time two-point correlator of a generic composite operator $\mathcal{O}_{\lambda}$ in the boundary theory using this deformed bulk spacetime \cite{Witten}. The correlator is defined as
\be\label{2point}
G^{\mathcal{O}_{\lambda}}(\omega,\varphi, \varphi')=-i\int^{\infty}_{0} dt e^{-i\omega t} \langle[\mathcal{O}_{\lambda}(t, \varphi),\mathcal{O}_{\lambda}(0, \varphi')]\rangle,
\ee
where $\lambda$ is the bulk field dual to $\mathcal{O}_{\lambda}$. Note that the two-point function poles encode the spectrum of states (including bound states and resonances) which could be created/annihilated by $\mathcal{O}_{\lambda}$. From this view point, it parallels the energy levels in the single-body Anderson Hamiltonian problem of \cite{Ault}, where the use of level statistics was advocated. 

Our static bulk geometry is asymptotically global AdS$_3$
 \be\label{bkg}
ds^2=-\frac{e^{4\Phi(\rho,\varphi)}}{\cos^2{\!\rho}}d\tau^2+\frac{e^{4\Psi(\rho,\varphi)}}{\cos^2\!\rho}(d\rho^2+\sin^2\!\rho d\varphi^2),  
\ee
where $\rho\in [0,\frac{\pi}{2}]$, $\varphi \in [0,2\pi]$. The AdS radius is set to one. The AdS$_{3}$ asymptotic boundary condition is implemented by requiring $\Phi, \Psi_{\rho\rightarrow\frac{\pi}{2}}\sim \Phi^{(b)}(\varphi), \Psi^{(b)}(\varphi)$, thus allowing for non-trivial boundary metrics. The Lagrangian density for the bulk field $\lambda$ is 
\be\label{LLambda}
\mathcal{L}=-\frac{1}{2}(\partial\lambda)^2-\frac{1}{2}m_{\lambda}^2\lambda^2+\mathcal{L}_{int},
\ee
where $\lambda$ is taken to be dual to a relevant deformation. We will only need the quadratic part of \eqref{LLambda}. Using the AdS/CFT dictionary, this setup is expected to capture aspects of a disordered large-N gauge theory defined on a 1+1-dimensional space with general metric. The spatial direction has a circular topology. We call the limit where $\Phi=\Psi=0$ ``clean''. 

In order to remain general, we avoid limiting ourselves to any particular AdS/CMT embedding of the disordered boundary theory. Our attitude will be that although the background \eqref{bkg} solves AdS$_3$ gravity coupled to matter equations of motion, nature of the matter source is left unspecified. Nevertheless, we demand that the bulk spacetime satisfies the Null Energy Condition (NEC) and is free of singularities.\footnote{Once temperature is introduced, a horizon boundary condition will replace this somewhat strong condition of no singularity.} This is done by checking if $R_{\alpha\beta}\zeta^{\alpha}\zeta^{\beta}\ge 0$ everywhere in the bulk for an arbitrary null vector $\zeta^{\alpha}$ \cite{Gubser}. This space of geometries could be called potentially realizable backgrounds (PRB). The important task of complete characterization of the PRB is left for future explorations. To begin, we will be content to study a class of backgrounds, where only $\Phi$ is turned on. For this class, equation of motion for the probe field $\lambda$ is the following partial differential equation (PDE)
\be\label{eom}
\sin^2{\!\rho}\partial^2_{\rho}\psi+\partial^2_{\varphi}\psi+[1+m^2_{\lambda}-\frac{4m^2_{\lambda}+3}{4\cos^2\!\rho}+\omega^{2}e^{-4\Phi}\sin^2{\!\rho}-\mathcal{K}(\Phi)]\psi=0,
\ee
where $\lambda=\tan^{-1/2}{\!\rho}~e^{-\Phi}\psi e^{-i\omega \tau}$ and 
\begin{eqnarray}
\mathcal{K}(\Phi)&=&\sin^2\!\rho[\partial^2_{\rho}\Phi+(\partial_{\rho}\Phi)^2] +\tan\!\rho~\partial_{\rho}\Phi\\&&\nonumber+[\partial^2_{\varphi}\Phi+(\partial_{\varphi}\Phi)^2].
\end{eqnarray}
The problem of extracting poles of the two-point correlator \eqref{2point} is the same as solving \eqref{eom} subject to boundary conditions as a PDE eigenvalue problem. To tackle the singular PDE in \eqref{eom}, only numerical techniques are available. In the simulations, varying $m_{\lambda}$ did not cause any qualitative change in the results reported here. Throughout this paper, we set $m_{\lambda}^2=-0.75$. With this choice of mass, there are two consistent quantizations of the bulk. We pick the one in which conformal dimension $\Delta_{\lambda}$ of the operator $\mathcal{O}_{\lambda}$ is $3/2$. The imposed boundary conditions are regularity of $\lambda$ at the origin of the global coordinates and vanishing of $\psi$ at the boundary. A family of backgrounds \eqref{bkg} 
\be\label{ss}
e^{4\Phi}=1+f(\rho)\sin^2{\!\varphi},\quad \Psi=0,
\ee
where $f(\rho)$ is regular at $\rho=\frac{\pi}{2}$ and $f(\rho)\sim\rho^2$ near $\rho=0$ to ensure the absence of singularities, contains spaces satisfying the NEC. It is necessary that $f(\rho)$ has no extremum.   

Before proceeding, let us briefly define the NNSD in the context of RMT \cite{RMT}. To meaningfully compare the fluctuation properties of two sequences of data, or different segments of a given sequence\footnote{This could be any sequence: energy levels of complex nuclei or spectrum of the operator in \eqref{eom}, for instance.}, it is necessary to first \emph{unfold} the sequence. This ensures that the non-universal differences, reflected in a non-constant average level density are removed. The unfolding procedure sets the average level density to one. In practice, to unfold a given spectrum one needs to numerically fit the level density histogram and then $x_{i}=\int^{\epsilon_{i}}_{\epsilon_b}\mathfrak{L}(\epsilon)d\epsilon$, where $\{x_i\}$ form the unfolded sequence, $\epsilon_b$ marks the beginning of the sequence, and $\mathfrak{L}$ denotes the fitted local level density.

As stated before, the order-parameter of interest to us is the NNSD of the eigenvalue sequence of \eqref{eom}, defined in the following. Take $\{\epsilon_1,\cdots,\epsilon_m\}$ to be a sequence of $m$ numbers, say eigenvalues of a large random matrix. Denote the corresponding unfolded sequence defined above, as $\{x_{1},\cdots, x_{m}\}$, where the sequence has further been sorted in an ascending order. Now form $s_{i}=(x_{i+1}-x_{i})/\langle s\rangle$, where the average is over the entire sequence $\{s_{i} \}$. The histogram associated to the set $\{s_{i}\}$ is the NNSD. Before delving into the numerics, a comment is in order. The NNSD is usually calculated for a data set collected from many realizations of the quenched disorder. Within the approach here, it is not clear that averaging over an ensemble of bulk geometries is the same as disorder-average in the boundary theory. We only focus on individual realizations. We hope to come back to this in the future.  

Powerful spectral techniques are utilized to numerically compute the spectrum of \eqref{eom} accurately \cite{NPDE}. We Chebyshev expand the wave function $\psi$ in the radial direction. Along the spatial boundary direction, a Fourier expansion is performed. To completely discretize the PDE, we perform collocation at Chebyshev-Gauss-Lobatto points \cite{NPDE}. In the ``clean'' limit, poles of the two-point function \eqref{2point} are given by $\omega=\pm(\Delta_{\lambda}+2k+\ell)$, where $k$, $\ell=0, 1, \cdots$\cite{Bala}. We use eigenvalues of the clean system as a check on the performance of our code. We refer to the collection of $N_d$ lowest eigenvalues computed numerically, and sorted in the ascending manner as our ``data set/sequence''. It is empirically observed that about one half of the total number of eigenvalues, produced in a computer run, are fully resolved. 
Furthermore, to ensure the accuracy of the computed eigenvalues, all the analysis below is repeated with higher spectral resolution until convergence is observed. In practice, roundoff errors are the limiting factor for large simulations with high spectral resolution. The eigenvalues come in pairs, due to symmetry $\tau\rightarrow-\tau$. Without loss of generality, the positive branch is picked. 

\begin{figure}[tbh]
 \begin{center}
\includegraphics[width=8cm]{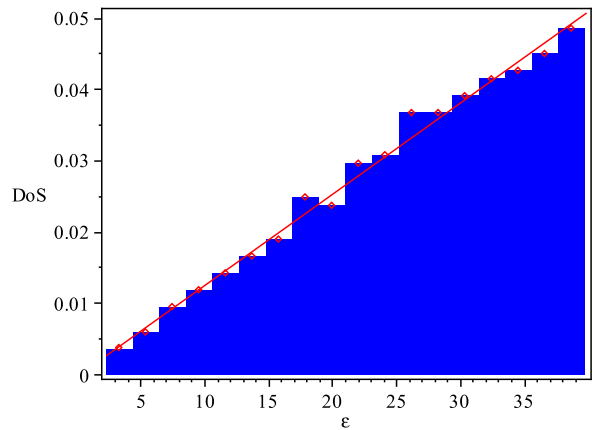}
\caption{(color online). Level density plot for $W=1.44$. Total number of participating eigenvalues is $N_d=406$.}
\label{fld}
\end{center}
\end{figure}

\begin{figure}[tbh]
\begin{center}
\includegraphics[width=8cm]{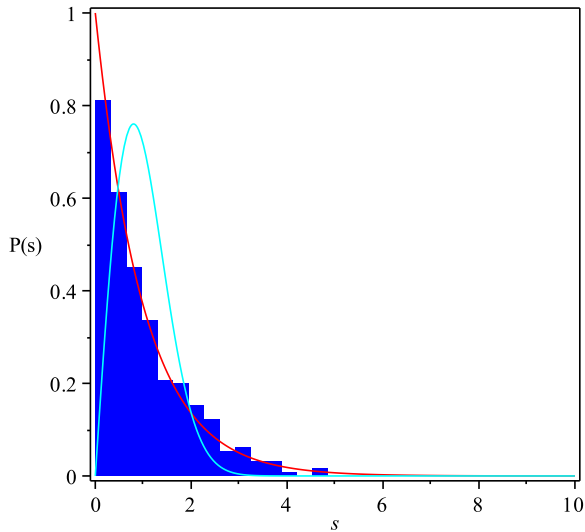}
\caption{(color online). The NNSD for 406 eigenvalues of the system \eqref{eom} with W=1.44. The cyan curve is the Wigner surmise for the GOE. The red curve is Poisson distribution $P(s)=\exp(-s)$. }
\label{NNSH}
\end{center}
\end{figure}

\begin{figure}[tbh]
\begin{center}
\includegraphics[width=8cm]{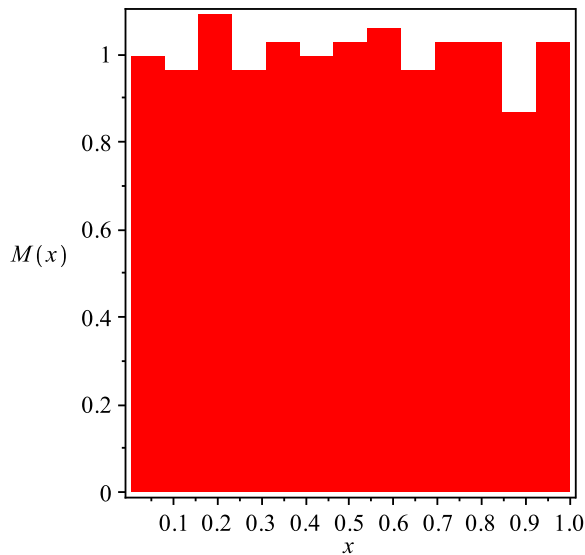}
\caption{(color online). Level density of the unfolded spectrum for $W=1.44$. It is nearly a constant (normalized to one) with the expected statistical fluctuations.}
\label{fig1}
\end{center}
\end{figure}

\begin{figure}[tbh]
\begin{center}
\includegraphics[width=8cm]{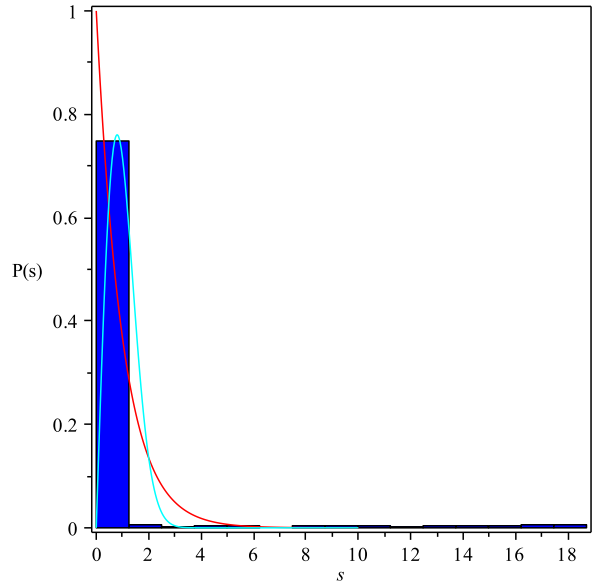}
\caption{(color online). The NNSD for $N_d=406$ eigenvalues of the system \eqref{eom} with $W=0.009$. The cyan curve is the Wigner surmise for the GOE. The red curve represents Poisson distribution $P(s)=\exp(-s)$.}
\label{NNSH0}
\end{center}
\end{figure}

\begin{figure}[th]
\begin{center}
\includegraphics[width=10cm]{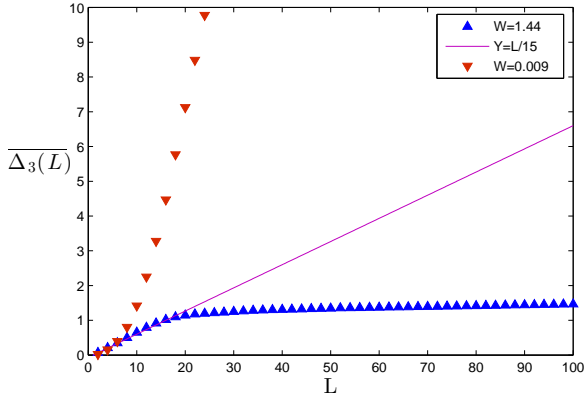}
\caption{(color online). The blue and red diamonds represent $\overline{\Delta_3(L)}$ computed from the data set of $N_d=406$ eigenvalues for $W=1.44$ and $W=0.009$, respectively. The line $y=L/15$ indicates the expected behavior of the statistic if the sequence is Poisson distributed.}
\label{D3data}
\end{center}
\end{figure}

Now we are prepared to present our numerical results. In order to illustrate the central idea  proposed in this paper, we study a simple model of the form \eqref{ss}, parametrized as $f(\rho)=W\rho^2(1+\rho^2)$, satisfying the NEC. The parameter $W>0$ will control the ``strength'' of the deformation $\Phi$. Note that $W$ can not be gauged-away. 
The use of Statistics is justified due to large number of poles involved in the analysis. We give sample values of $W$ which correspond to Poisson and non-Poisson phases of the NNSD of the pole spectra. A more systematic search on the space of PRB will be reported elsewhere. A common method of testing a statistical hypothesis about a given data set is to apply a $\chi$-squared test. For a given value of $W$, we perform the $\chi$-squared test to compare the numerically computed NNSD of the eigenvalues of \eqref{eom} against Poisson distribution. The result is summarized in the so called p-${value}$ index. The index determines whether there is statistical evidence against the null hypothesis.\footnote{The null hypothesis here is ``for a given value of $W$, the corresponding NNSD is Poisson''.} If the associated p-$value$ of the $\chi$-squared test is larger than a (commonly accepted) threshold value of 0.05, data is considered consistent with the distribution. The level density plot for $W=1.44$ appears in Fig.~\ref{fld}. There is no universal recipe for choosing the ``right'' number of bins, although there exist recommended values for the bin numbers, if one has prior knowledge about the distribution. We avoid any bias in choosing the number of bins. A cautionary note is that one should not attempt to fit the histogram \emph{too well}. In particular, a fit involving high degree polynomials is undesirable. We apply linear regression to the level density histogram. The fitting line is also overlaid on the level density histogram in Fig.~\ref{fld}.
The corresponding histogram for the unfolded data is given in Fig.~\ref{fig1}. 
The NNSD for $W=1.44$ is presented in Fig.~\ref{NNSH}. The corresponding p-$value$ is computed to be approximately 0.93, which indicates that our data set is consistent with the null hypothesis. 
The NNSD histogram for $W=0.009$ is shown in Fig.~\ref{NNSH0} with a p-$value\approx$0, which indicates a non-Poisson NNSD. 
In the following, we turn to $W=1.44$ case and perform yet another statistical tests to build confidence that the NNSD is indeed Poisson. In the context of RMT, Dyson and Mehta suggested several statistics for studying the spectrum of random matrices, the more useful one being $\Delta_{3}$, the so-called spectral rigidity \cite{Dyson}. This statistic is sensitive to the long-range correlations in the spectrum. It is defined as follows 
\be
\overline{\Delta_{3}(L)}=\frac{1}{L}\langle \textrm{min} \int^{x_i+L}_{x_i} [M(x)-Ax-B]^2dx\rangle=\langle \Delta^{i}_3(L)\rangle,
\ee 
where the average is over ``$i$'', the position of the beginning of the sequence, and $M(x)$ is the integrated level density for the unfolded sequence. The symbol ``min'' indicates that $\Delta_3(L)$ is calculated for the values $A$ and $B$ for which the right hand side in minimized.
It is well known that for a Poisson distribution of levels, for large values of $L$, the spectral rigidity has a linear behavior with a constant slope 
\be
\overline{\Delta^{i}_{3}(L)}\rightarrow \frac{1}{15}L.
\ee
Applying this to our data set of eigenvalues, the plot in Fig.~\ref{D3data} is obtained. 
As is clearly observed, the case $W=1.44$ corresponds to a Poisson sequence, while $W=0.009$ is non-Poisson. Within the approach taken here, it might not be meaningful to determine the critical value $W_c$ where the transition happens, since it is not immediately clear how to connect $W$ to disorder in the boundary field theory. All that is known is that $W=0$ is the ``clean'' limit and $W\neq0$ should correspond to ``disordered''. The observed saturation of the linear behavior is well understood from the work of M. V. Berry \cite{Berry}. We also applied our analysis to the case where the background \eqref{bkg} has a more general $\varphi$-dependence. Moreover, we studied (unsystematically) the case where $f(\rho)\rightarrow 0$ as the boundary was approached. For these cases, the results were qualitatively similar to the ones reported in this paper.

{\it Discussion}-

We have demonstrated that at least sub-sectors of the spectrum of a disordered large-N gauge theory with gravity dual can become entirely uncorrelated/random. It would be interesting to further characterize the transition. 
It is also crucial to investigate how generic our findings are. 
Important steps in understanding Anderson localization in a many-body setting were reported in \cite{Ault2}, \cite{Huse}. 
The resemblance we find here with the localized phase of the Anderson Hamiltonian is striking, although much needs to be done before one could conclusively claim that the NNSD transition observed here  is connected in any way to a many-body version of the AL transition. The AL transition will have signatures in the transport channel as well which are currently under investigation \cite{AllanOmid}. Another important question left for the future is to find out whether change in the NNDS is correlated with other signatures of the AL transition observed in transport, for instance. 

The issue of dependence on the realization needs further clarification. Perhaps the right approach is to average over many spectra, each generated by a particular realization of the disordered system.
It would be interesting to generalize our results to higher-dimensional settings. Simulations involving a horizon in the bulk will teach us about statistics of resonances in the thermal plasma of a disordered gauge theory. The corresponding results will be reported elsewhere. 

Another future direction is to specialize to a particular embedding of the disordered boundary theory, where one solves for the back-reacted gravity background and computes the critical disorder strength at which transition in the NNSD takes place.

\begin{acknowledgements}

O.S. is grateful to Allan Adams for fruitful discussions, comments and collaboration on related topics. Special thanks to Sean Hartnoll for useful discussions, suggestions and comments. O.S. would like to thank Dionysios Anninos, Mohammad Edalati, Tarun Grover and John McGreevy for comments and Alex Dahlen, Shannon McCurdy, Valdimir Rosenhaus and Kevin Schaeffer for helping with the manuscript. O.S. is supported by the Berkeley Center for Theoretical Physics, department of physics at UC Berkeley and in part by DOE, under contract DE-AC02-05CH11231.
\end{acknowledgements}


\begin{thebibliography}{9}

\bibitem{Anderson} 
  P.~W.~Anderson,
  ``Absence of Diffusion in Certain Random Lattices,''
  Phys.\ Rev.\  {\bf 109}, 1492 (1958).

\bibitem{Malda}
  J.~M.~Maldacena,
  ``The Large N limit of superconformal field theories and supergravity,''
  Adv.\ Theor.\ Math.\ Phys.\  {\bf 2} (1998) 231
   [Int.\ J.\ Theor.\ Phys.\  {\bf 38} (1999) 1113]
  [hep-th/9711200].
  

 
  \bibitem{Sean}
  S.~A.~Hartnoll,
  ``Lectures on holographic methods for condensed matter physics,''
  Class.\ Quant.\ Grav.\  {\bf 26}, 224002 (2009)
  [arXiv:0903.3246 [hep-th]].
  

\bibitem{John} 
  J.~McGreevy,
  ``Holographic duality with a view toward many-body physics,''
  Adv.\ High Energy Phys.\  {\bf 2010}, 723105 (2010)
  [arXiv:0909.0518 [hep-th]].


\bibitem{SeanChris} 
  S.~A.~Hartnoll and C.~P.~Herzog,
  ``Impure AdS/CFT correspondence,''
  Phys.\ Rev.\ D {\bf 77}, 106009 (2008)
  [arXiv:0801.1693 [hep-th]].
  

\bibitem{Adams1} 
  A.~Adams and S.~Yaida,
  ``Disordered Holographic Systems I: Functional Renormalization,''
  arXiv:1102.2892 [hep-th].


\bibitem{Adams2} 
  A.~Adams and S.~Yaida,
  ``Disordered Holographic Systems II: Marginal Relevance of Imperfection,''
  arXiv:1201.6366 [hep-th].

\bibitem{Dio1} 

  D.~Anninos, T.~Anous, J.~Barandes, F.~Denef and B.~Gaasbeek,
  ``Hot Halos and Galactic Glasses,''
  JHEP {\bf 1201}, 003 (2012)
  [arXiv:1108.5821 [hep-th]].

\bibitem{Dio2} 
  D.~Anninos, T.~Anous, F.~Denef, G.~Konstantinidis and E.~Shaghoulian,
  ``Supergoop Dynamics,''
  arXiv:1205.1060 [hep-th].

  \bibitem{Taka} 
  M.~Fujita, Y.~Hikida, S.~Ryu and T.~Takayanagi,
  ``Disordered Systems and the Replica Method in AdS/CFT,''
  JHEP {\bf 0812}, 065 (2008)
  [arXiv:0810.5394 [hep-th]].

  
\bibitem{Ault}
B. L. Altshuler, I. Kh. Zharekeshev, S. A. Kotochigova, and B. I. Shklovskii, Zh. Eksp. Teor. Fiz. {\bf 94}, 343 (1988)[Sov. \ Phys. JETP {\bf 67}, 625 (1988)].
  
\bibitem{Imry}

U.Sivan and Y. Imry, Phys. Rev. B {\bf 35}, 6074 (1987).

\bibitem{Izra}
F. M. Izrailev, Phys. \ Rep. {\bf 129}, 299 (1990).

\bibitem{Shk}
B. I. Shklovskii, B. Shapiro, B. R. Sears, P. Lambrianides, and H. B. Shore, Phys. Rev. B {\bf 47}, 11487 (1993). 

\bibitem{Witten} 

  E.~Witten,
  ``Anti-de Sitter space and holography,''
  Adv.\ Theor.\ Math.\ Phys.\  {\bf 2}, 253 (1998)
  [hep-th/9802150];  S.~S.~Gubser, I.~R.~Klebanov and A.~M.~Polyakov,
  ``Gauge theory correlators from noncritical string theory,''
  Phys.\ Lett.\ B {\bf 428}, 105 (1998)
  [hep-th/9802109].

\bibitem{Gubser} 

  D.~Z.~Freedman, S.~S.~Gubser, K.~Pilch and N.~P.~Warner,
  ``Renormalization group flows from holography supersymmetry and a c theorem,''
  Adv.\ Theor.\ Math.\ Phys.\  {\bf 3}, 363 (1999)
  [hep-th/9904017].

\bibitem{RMT}

M. L. Mehta, {\it Random Matrices} (Academic, New York, 1991).

 \bibitem{NPDE}
 
 David Kopriva, {\it Implementing Spectral Methods for Partial Differential Equations: Algorithms for Scientists and Engineers}, (springer, 2009).

\bibitem{Bala} 
 
  V.~Balasubramanian, P.~Kraus and A.~E.~Lawrence,
  ``Bulk versus boundary dynamics in anti-de Sitter space-time,''
  Phys.\ Rev.\ D {\bf 59}, 046003 (1999)
  [hep-th/9805171].
 
  
\bibitem{Dyson}

F. ~J.~ Dyson, L.~M.~ Mehta, \ J.\ Math. \ Phys. {\bf 4}, 701 (1963).


\bibitem{Berry}

M.~V.~Berry, Semiclassical  theory of spectral spectral rigidity, Proc.\ R. \ Soc. \ Lond.\  A {\bf 400}, 229 (1985).


\bibitem{Ault2}

D. M. Basko, I. L. Aleiner, and B. L. Altshuler, Ann. Phys. {\bf 321}, 1126 (2006).

\bibitem{Huse}

Arijeet Pal, David Huse, Phys. Rev B {\bf 82}, 174411 (2010).

\bibitem{AllanOmid}

Allan Adams, Omid Saremi, Sho Yaida, work in preparation. 

  
\end{thebibliography}
\end{document}